# Critical Magnetic Field Ratio of Anisotropic Magnetic Superconductors

A.Changjan[1,2,3] and P.Udomsamuthirun[1,2]

[1] Prasarnmitr Physics Research Unit, Department of Physics, Faculty of Science, Srinakharinwirot University Bangkok 10110, Thailand.
e-mail:udomsamut55@yahoo.com
[2] Thailand Center of Excellence in Physics, Si Ayutthaya Road, Bangkok 10400, Thailand.
[3] Department of Mathematics and Basic science, Faculty of Science and Technology, Phathumwan Institute of Technology, Bangkok 10330, Thailand.

**Abstract**

The upper critical field, the lower critical field and the critical magnetic field ratio of anisotropic magnetic superconductors are calculated by Ginzburg-Landau theory analytically. The effect of the Ginzburg-Landau parameter($\kappa_0$), magnetic susceptibility($\chi$) and magnetic-to-anisotropic parameter ratio($\theta$) on the critical field ratio are considered. We find that the value of critical magnetic field ratio with $\chi < 0$, higher $\kappa_0$, and higher $\theta$ case are shown the higher value. And, the diamagnetic superconductors with highly anisotropic and the ferromagnetic superconductors are shown the highest and the lowest, respectively.

**1. Introduction**

The Ginzburg-Landau(GL) theory is widely used in framework for describing metallic and magnetic superconductors in magnetic field [1,2,3]. For type I superconductors, the magnetic response is diamagnetic. It shows total exclusion of flux in low magnetic field. For type II superconductors, there are quantized flux penetrates the superconductors in high magnetic field. The magnetic response of these material can be different from type I. The properties of type-II superconductors at low and high applied magnetic field ,close to lower and upper critical field was studied by many researchers . Hampshire[4,5] studied the magnetic superconductors by GL theory including the spatial variation and nonlinear magnetic response of magnetic ions in-field. The analytic formula of upper critical field is shown. Askerzade[6] studied two-band GL theory and apply to determine the temperature dependence of lower ,upper and thermodynamic critical field for non-magnetic superconductors. Askerzade[7] , Udomsamuthirun et al.[8],Min-xia and Zi-Zhao[9] studied the upper critical field of anisotropy two-band superconductors by GL theory.

In this paper, we calculate analytically the upper critical field, the lower critical field and the critical magnetic field ratio of the anisotropic magnetic superconductor by Ginzburg-Landau(GL) theory that the Gibbs free energy of magnetic superconductors of Hampshire[4,5] are used.

**2. Model and Calculation**

According to the Ginzburg and Landau[1] theory, the Helmholtz free energy of non-magnetic superconductors is of the form

$$F_s(H,T) = F_N + \alpha|\psi|^2 + \frac{1}{2}\beta|\psi|^4 + \frac{1}{2m}\left|\left(-i\hbar\vec{\nabla} - 2e\vec{A}\right)\psi\right|^2 + \int H_s \, dB \qquad (1)$$

Here, $F_N$ ,and $F_s$ are the Helmholtz free energy in the normal state and superconducting state,

$B$ is the net field in the superconductivity region,

$\psi$ is the order parameter and $\left(|\psi|^2\right)$ is proportional to the density of carriers,

$m$ denotes the effective mass of the carriers,

the coefficient $\alpha$ depends linearly on the temperature, while coefficient $\beta$ is independent of temperature.

The fourth term accounts for the kinetic energy of the carriers and the lasted term accounts for the energy stored in the local magnetic fields.

Hampshire[4,5] proposed the Gibbs's free energy of magnetic superconductors by assuming that $G_s(B,T) = F_N(B,T) - \mu_0 HM$ ,

$H_s = \dfrac{B}{\mu_0} - M_{ions}$, $M = M_{sc} + M_{ions}$ and $M_{ions} = \chi H_s$. Therefore Gibbs free energy of magnetic superconductors can be written as

$$G_s(B,T) = F_N + \alpha|\psi|^2 + \dfrac{1}{2}\beta|\psi|^4 + \dfrac{1}{2m}\left|\left(-i\hbar\bar{\nabla} - 2e\bar{A}\right)\psi\right|^2 + \int(B - \mu_0 M_{ions})\dfrac{dB}{\mu_0} - (B - \mu_0 M)M \qquad (2)$$

For small change in B-field, a series in B is introduced

$$\gamma_0 + \gamma_1 B + \gamma_2 \dfrac{B^2}{2\mu_0} = \int(B - \mu_0 M_{ions})\dfrac{dB}{\mu_0} - (B - \mu_0 M)M_{sc} - (B - \mu_0 M)M_{ions}$$

Here $\gamma_0, \gamma_1$ and $\gamma_2$ are coefficient parameters. We can get the $\gamma_2$ by differentiating above equation twice and taking $M_{sc}$ to be small that

$$\gamma_2 = 1 - \mu_0 \dfrac{dM_{ions}}{dB} - \mu_0 \dfrac{d^2}{dB^2}\left[(B - \mu_0 M_{ions})M_{ions}\right]$$

Here $\gamma_2 = 1$ for non-magnetic superconductors. The coefficient parameter $\gamma_1$ depends on the gradient of field and $\gamma_1 = 0$ for the uniform applied field.

Finally, the Gibbs's free energy of magnetic superconductors can be written as

$$G_s(B,T) = F_N + \alpha|\psi|^2 + \dfrac{1}{2}\beta|\psi|^4 + \dfrac{1}{2m}\left|\left(-i\hbar\bar{\nabla} - 2e\bar{A}\right)\psi\right|^2 + \gamma_0 + \gamma_1 + \gamma_2 \dfrac{B^2}{2\mu_0} \qquad (3)$$

When the magnetic superconductors place in magnetic field, there are the field produced by ions $(\mu_0 M_{ions})$, the applied field $(\mu_0 H)$ and the field produced by the carriers $(\mu_0 M_{sc})$ that can produce relationship between field as

$$B = \mu_0 H + \mu_0 M_{sc} + \mu_0 M_{ions} \qquad (4)$$
$$M_{ions} = \chi H_{c_2}(T) + \chi'(H - M_{sc} - H_{c_2}) \qquad (5)$$

So that
$$B = \mu_0(\chi - \chi')H_{c_2} + \mu_0(1 + \chi')(H + M_{sc}) \qquad (6)$$

where the differential susceptibility $(\chi')$ is $\left(\dfrac{\partial M_{ions}}{\partial H}\right)_{T, H=H_{c_2}}$ and susceptibility $(\chi)$ is $\left(\dfrac{M_{ions}}{H}\right)_{T, H=H_{c_2}}$.

By comparing Ginzburg-Landau theory to BCS theory, we can get $\psi \sim \Delta$ that $\Delta$ is gap function from BCS theory and $\psi$ is the order parameter from GL theory. And, the anisotropic gap function[10] can be written as

$$\Delta(\hat{k},T) \sim f(\hat{k})\Delta(T)$$

where $f(\hat{k})$ is the anisotropic function. Similarity, we can write anisotropic order parameter [8]

$$\psi(\hat{k},T) = f(\hat{k})\psi(T) \qquad (7)$$

The anisotropy gap function has been propose by Haas and Maki[11] : $f(\delta) = \dfrac{1 + a\cos^2 \delta}{1 + a}$ and by Posazhennikova et al.[12] : $f(\delta) = \dfrac{1}{\sqrt{1 + a\cos^2 \delta}}$, here $\delta$ is the polar angle and $a$ is anisotropy parameter. In case of the symmetry order parameter $f(\hat{k}) = 1$.

Minimising $G_s$ of Eq.(3) with respect to $\psi^*$ and $\bar{A}$, here we use anisotropic order parameter in Eq.(7), they lead to the 1$^{st}$ and 2$^{nd}$ Ginzburg-Landau equation, which now include the effect of anisotropic function and magnetic ions as

$$\alpha\psi\langle f^2(\hat{k})\rangle + \beta|\psi|^2\psi\langle f^4(\hat{k})\rangle + \dfrac{1}{2m}\left(-i\hbar\bar{\nabla} - 2e\bar{A}\right)^2\psi\langle f^2(\hat{k})\rangle = 0 \qquad (8)$$

$$\left(\dfrac{\gamma_1}{B} + \dfrac{\gamma_2}{\mu_0}\right)\mu_0\bar{\nabla}\times(1+\chi')M_{sc} = \left[-\dfrac{ie\hbar}{m}(\psi^*\bar{\nabla}\psi - \psi\bar{\nabla}\psi^*) - \dfrac{4e^2\bar{A}|\psi|^2}{m}\right]\langle f^2(\hat{k})\rangle$$

(9)

Where, $\bar{\nabla}\times M_{sc} = J_{sc}$ is the supercurrent density and $<...>$ is averaged over Fermi surface.

Eq.(8) and Eq.(9) can be reduced to the isotropic magnetic superconductors by setting $\langle...\rangle$ to 1 and $\gamma_1 = 0$.

For the uniform applied field, the upper critical field can be found from the linearised 1$^{st}$ Ginzburg-Landau equation, Eq.(8). Then, we get

$$\xi^2(T) = \dfrac{\hbar^2}{2m|\alpha|} \qquad (10)$$

$$\mu_0 H_{c_2} = \dfrac{\phi_0}{2\pi\xi^2(T)(1+\chi)} = \dfrac{\mu_0 H_{c_2}(\chi=0,T)}{(1+\chi)} \qquad (11)$$

Here $\xi$ is the coherence length, $\mu_0 H_{c_2}$ is the upper critical field that magnetic flux quantum $(\phi_0)$ is $\dfrac{\pi\hbar}{e}$ and $H_{c_2}(\chi=0,T)$ is upper critical field of nonmagnetic superconductors.

Eqs.(10-11) show that the coherence length is unchanged by the presence of the magnetic ions and anisotropic function. And, the presence of the ions reduces the upper critical field strength by a factor $(1+\chi)$, but the effect of anisotropic function is not found.

For lower critical field's calculation, we choose a flux line in cylindrical coordinate $\vec{B} = B(r)\hat{z}$ where $B(r)$ has its maximum value at core and tends to zero at large radial. A vector potential in cylindrical coordinate is chosen as

$$\vec{A} = [\mu_0(\chi - \chi')H_{c_2}r + \mu_0(1+\chi')(H + M_{sc})r]\hat{\theta} = A(r)\hat{\theta} \qquad (12)$$

From the solution for the wave function $\psi(T)$ provided by

$$\psi(T) = |\psi(T)|e^{i\phi} \qquad (13)$$

where $\phi$ is the phase of order parameter.

By using the relevant Maxwell's equation, $\vec{j} = \dfrac{1}{\mu_0}\vec{\nabla}\times\vec{B}$ (for the magnetostatic case, $\dfrac{\partial \vec{D}}{\partial t} = 0$) and 2$^{nd}$ Ginzburg-Landau equation in uniform field, the vector potential and magnetic flux quantum are $\vec{A} = \dfrac{\hbar}{2e}\vec{\nabla}\phi$ and $\Phi = \dfrac{2\pi\hbar}{2e}$. The equation for the vector potential is taken the form

$$\vec{\nabla}\times\vec{\nabla}\times\vec{B} = -\dfrac{\mu_0}{\gamma_{2(1+\chi')}}\cdot\dfrac{4e^2|\alpha|}{m\beta}(\vec{\nabla}\times\vec{A})\langle f^2(\hat{k})\rangle = -\dfrac{\mu_0}{\gamma_{2(1+\chi')}}\cdot\dfrac{4e^2|\alpha|\vec{B}}{m\beta}\langle f^2(\hat{k})\rangle$$

Let our sample has a surface perpendicular to the x-axis and the external magnetic field is $\vec{B} = B_0\hat{z}$, so the internal magnetic field should be the form $\vec{b} = b(x)\hat{z}$. Then, the London equation is of the form $\dfrac{d^2 b(x)}{dx^2} - \dfrac{1}{\lambda^2}b(x) = 0$. Here, $\lambda$ is the London penetration depth of anisotropic magnetic superconductors

$$\lambda = \left(\dfrac{\gamma_2(1+\chi')m\beta}{\mu_0(4e^2)\langle f^2(\hat{k})\rangle|\alpha|}\right)^{\frac{1}{2}} \qquad (14)$$

The lower critical field can be obtained as

$$B_{c_1} = \dfrac{\hbar}{4e\lambda^2}\ln\kappa \qquad (15)$$

Where $\kappa$ is Ginzburg-Landau parameter ; $\kappa = \frac{\lambda}{\xi}$. The lower critical field Eq.(15) is the same form of Abrikosov[2] but it has the difference in $\kappa$ that depend on $\lambda$ in Eq.(14).

We introduce a dimensionless parameter, the critical magnetic field ratio as

$$\eta = \frac{B_{c_2}}{B_{c_1}} \quad (16)$$

Substitution Eq.(11) and Eq.(15) in Eq.(16), we can get

$$\eta = \frac{2\kappa^2}{(1+\chi)\ln\kappa}$$
$$= \frac{4\theta\kappa_0^2}{(1+\chi)\ln\theta\kappa_0^2} \quad (17)$$

In case $\kappa \gg 1$, the approximated critical magnetic field ratio is

$$\eta \approx \frac{2\theta\kappa_0^2}{(1+\chi)(\sqrt{\theta}\,\kappa_0 - 1)} \quad (18)$$

Here $\kappa_0$ is the isotropic non-magnetic Ginzburg-Landau parameter; $\kappa_0 = \frac{\lambda_0}{\xi_0}$ and $\kappa^2 = \theta\kappa_0^2$ that $\lambda_0$ and $\xi_0$ are the penetration depth and coherence length of Abrikosov[2]. The $\theta = \frac{\gamma_2(1+\chi')}{\langle f^2(\hat{k})\rangle}$ is the magnetic-to-anisotropic parameter ratio. For the pure superconductors that is the isotropic non-magnetic superconductors, Eq.(17) can be reduced to $\eta = \frac{2\kappa_0^2}{\ln\kappa_0}$ that agreed with the Ginzburg-Landau relation.

**3. Results and Discussions**

The effect of $\kappa_0, \chi$ and $\theta$ on the critical magnetic field ratio($\eta$) are shown in Figure 1. Here, we use the $\kappa_0 = 57$, and $\kappa_0 = 118$ that are of $Y123$ [13] and $Hg1223$ [14] superconductor, respectively. We consider in 3 cases the non-magnetic superconductor case($\chi = 0$), the diamagnetic case ($\chi < 0$) and ferromagnetic case($\chi > 0$). The $\eta$ must be the positive values so the value of $\gamma_2$ is between 0 to 1. The $\theta$ represents the ratio of magnetic parameter to anisotropic parameter. And $\theta = 1$, $\theta > 1$, $\theta < 1$ are the isotropic non-magnetic case, the highly anisotropic case and highly magnetic case, respectively. We find that the case $\chi < 0$, the higher critical field ratio, the higher $\kappa_0$ with higher $\theta$ is found. The diamagnetic superconductors($\chi < 0$) with highly anisotropic case ($\theta > 1$) shows the highest value and the ferromagnetic superconductor ($\chi>0$) with highly magnetic case ($\theta<1$) shows the lowest value . We also find that in the ferromagnetic superconductors, the difference between the upper critical field and the lower critical field is smaller than diamagnetic superconductors.

Yadav and Paulose[15] measured the lower and upper critical magnetic field of $FeTe_{0.06}Se_{0.40}$ superconductor. They found the upper critical magnetic field about 65 T for mid-point curve , the lower critical magnetic field about 82 Oe or $1.29 x 10^{-4}$ T ,and the magnetic susceptibility $\chi=-1.35 x 10^{-2}$ emu/gm. From Eq.(16), we can estimate the critical magnetic field ratio of this material as $\eta=5.04 x 10^5$. According to this $\eta$, and Eq.(17), the estimated Ginzburg-Landua parameter of this material should be between $\kappa_0 \cong 1337$ to $\kappa_0 \cong 109$ for $\theta=1-150$. Because $FeTe_{0.06}Se_{0.40}$ shows the anisotropic and magnetic property, we think that the $\kappa_0$ of material should be the same order Y123 material.

### 4. Conclusions

The upper critical, the lower critical field and the critical magnetic field ratio of anisotropic magnetic superconductors are calculated by Ginzburg-Landau theory analytically. The effect of the Ginzburg-Landau parameter, magnetic susceptibility and magnetic to anisotropic parameter ratio on the critical field ratio are considered. We find that the critical field ratio of $\chi<0$ case, with the higher $\kappa$ ,and higher $\theta$ ,the higher value of is the critical field ratio found. The diamagnetic superconductor with highly anisotropic case is shown the highest value and the ferromagnetic superconductor with highly magnetic is shown the lowest value of the critical magnetic field ratio. We can use the critical magnetic ratio to predicted the magnetic parameters of superconductors.

**Acknowledgements**

The authors would like to thank Professor Dr.Suthat Yoksan for the useful discussions. The financial support of the Srinakharinwirot University, Phathumwan Institute of Technology and ThEP center are acknowledged.

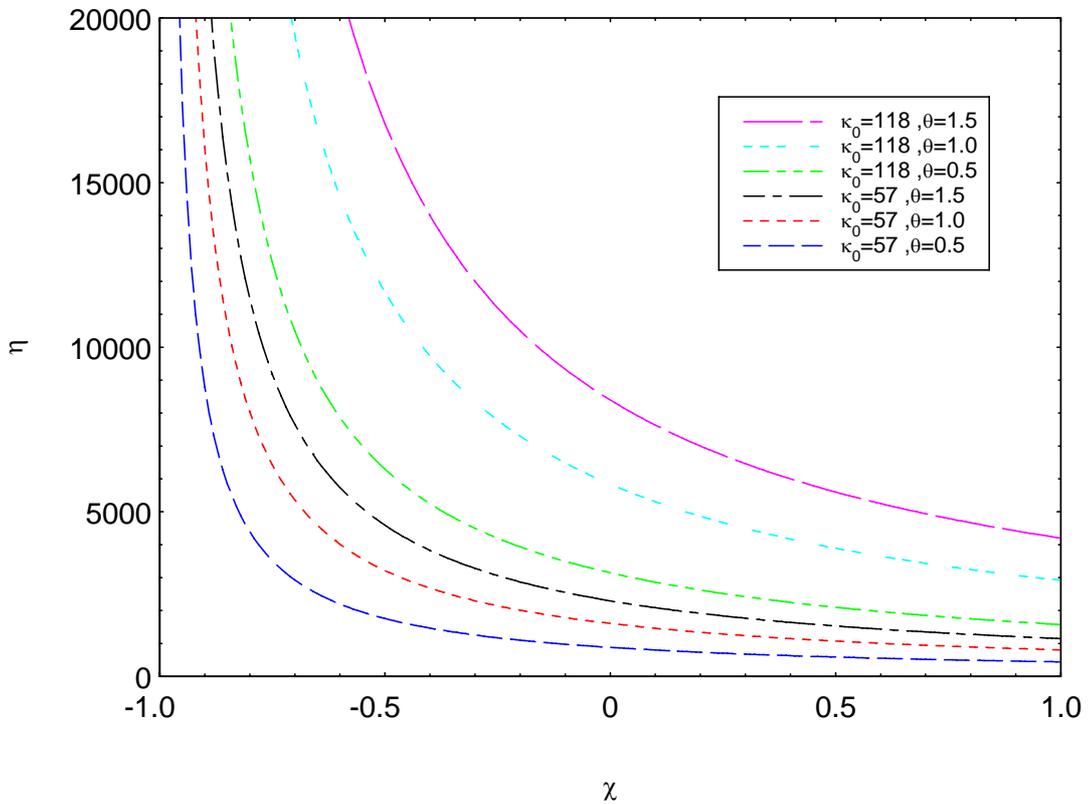

Figure 1. The critical magnetic field ratio versus the critical magnetic susceptibility of the anisotropic magnetic superconductors.